\documentclass[11pt,a4paper,naustrian,english]{article}
\usepackage[latin9]{inputenc}
\usepackage{color}
\usepackage{url}
\usepackage{amsmath}
\usepackage{amssymb}
\usepackage{graphicx}

\makeatletter

\newcommand{\lyxaddress}[1]{
\par {\raggedright #1
\vspace{1.4em}
\noindent\par}
}

\usepackage{latexsym}

\usepackage{babel}

\makeatother

\usepackage{babel}
\begin{document}

\title{\begin{flushright}  {\small UWThPh-2015-34 \,}  \bigskip \bigskip \bigskip \end{flushright}Ergostatting
and Thermostatting at a Fixed Point}

\author{Helmuth Hüffel$^{\,\,1}$ and Sa{\v{s}}a Iliji{\'{c}}$^{\,\,2}$ }

\maketitle

\lyxaddress{\begin{center}
$^{1}$$\,$ {\small Faculty of Physics, University of Vienna, Boltzmanngasse
5, A-1090 Wien }\smallskip{}
 \\
 $^{2}$$\,$ {\small University of Zagreb, Faculty of Electrical
Engineering and Computing, Department of Applied Physics, Unska 3,
HR-10000 Zagreb, Croatia} 
\par\end{center}}
\begin{abstract}
\noindent We propose a novel type of ergostats and thermostats for
molecular dynamics simulations. A general class of active particle
swarm models is considered, where any specific total energy (alternatively
any specific temperature) can be provided at a fixed point of the
evolution of the swarm. We identify the extended system feedback force
of the Nosé - Hoover thermostat with the ``internal energy'' $\mbox{variable}$
of active Brownian motion. 
\end{abstract}

\section{Introduction}

\global\long\def\uv{{\mathbf{e}}}
 \global\long\def\dotuv{\dot{\mathbf{e}}}

In this paper a novel type of ergostats and thermostats for molecular
dynamics simulations is proposed, which is derived from particle swarm
models. An assigned total energy or temperature can be provided at
fixed points of the evolution of the swarms.

Simulations in molecular dynamics are usually performed in the microcanonical
ensemble, where the number of particles, volume, and energy have constant
values. In experiments, however, it is the temperature which is controlled
instead of the energy. Several methods have been advanced for keeping
the temperature constant in molecular dynamics simulations. Popular
are deterministic thermostats like velocity rescaling \cite{velocity rescaling},
the Andersen thermostat \cite{andersen}, the Nosé\textendash{}Hoover
thermostat \cite{nose1,nose,nose progress,hoover,posch chaos} and
its generalizations \cite{kusnezov1,kusnezov2,jellinek,branka,branka2},
Nosé\textendash{}Hoover chains \cite{chains}, and the Berendsen thermostat
\cite{berendsen}. Gauss' principle of least constraint was utilized
by Evans, Hoover and collaborators to develop isokinetic \cite{isokinetic},
as well as isoenergetic (=ergostatic) thermostats \cite{isoenergetic}.
Dettmann and Morriss \cite{dettmann morris,dettmann} as well as Bond,
Laird, and Leimkuhler \cite{bond leimkuhler} discovered Hamiltonian
schemes for both the Nosé- Hoover and Gaussian thermostats. Another
setting arises for stochastic thermostats, which includes standard
Brownian (overdamped Langevin) and Langevin dynamics, as well as stochastic
thermostats of Nosé\textendash{}Hoover\textendash{}Langevin type \cite{gentle,slow modes}
and generalizations thereof \cite{generalized}. Stochastic velocity
rescaling which can be considered as Berendsen thermostat plus a stochastic
correction leading to canonical sampling was considered by \cite{bussi,bussi accurate,bussi 2,path integral langevin}.
For further discussion on the various thermostatting schemes, we refer
to the recent monographs \cite{klages book,leimkuhler book,hoover hoover book}.

Swarming - the collective, coherent, self organized motion of a large
number of organisms - is one of the most familiar and widespread biological
phenomena at the interface of physics and biology. Universal features
of swarming have been identified and diverse physical models of swarming
have been proposed (see the reviews \cite{toner,roman,vicsek}). 

Amongst these models there is a whole class, often referred to as
\textit{$\mbox{active}$} \cite{schimansky,schw und ebe}, which provides
a relevant tool for simulating complex systems (see the monographs
\cite{schweitzer book,ebeling book}). The notion active refers to
the property of particles to take up energy from their environment
and store it as so-called \textit{internal energy}. This then is followed
by the generation of an out-of equilibrium state of the system and
depending on the particular circumstances is implying self-propulsion,
alignment, attraction or repulsion of the particles. 

Recently a simple model for particle swarms was proposed \cite{paper,swarm},
which is the starting point for our present discussion. The model
is specified by a $(2dN+1)$-dimensional system of first order differential
equations for coordinates and momenta of $N$ active particles in
$d$ space dimensions coupled to internal energy. Such a nonlinear
system is not easily accessible with direct analytic procedures. Nonetheless,
precise predictions for the system's long time behavior can be made
in the case where all particles are attracted with harmonic forces.
We focus on the time evolution of macroscopic swarm variables, represented
by the total kinetic energy, total potential energy, virial and -
last but not least - internal energy. A closed four-dimensional system
of first order differential equations for the time evolution of these
macroscopic swarm variables can be obtained. In the long time limit
one finds a stable equilibrium configuration with fixed non-zero total
kinetic and potential energy, whereas internal energy$ $ and virial
are vanishing, see Fig.~\ref{fig:energiesfigs_h_ANH512.pdf} and
Fig.~\ref{fig:xifig_h_ANH512.pdf} below. Bifurcation analysis provides
us with conditions on parameters of the model for this to take place
\cite{paper,swarm}. \\
\\
It is intriguing to observe that in the equilibrium state with fixed
total kinetic energy the system effectively becomes thermostatted.
Thus a novel and original method of thermostatting at fixed points
of the evolution of particle swarms has been obtained. \textcolor{black}{The
usual Nosé-Hoover dynamics }\cite{nose1,nose,nose progress,hoover,posch chaos}\textcolor{black}{{}
has no} attractive fixed point, and in contrast to the Gaussian isokinetic
\cite{isokinetic} thermostat, no constraint needs to be implemented.
It is in the fixed point limit of the system's evolution that the
total kinetic energy becomes conserved. \\

In Section 2 we shortly review active multi-particle systems. Then
in Section 3 basic features of the fixed point method for thermostatting
as well as the related procedure for ergostatting are outlined. In
Section 4 prototype studies of the active ergostat and  thermostat
are given for a single particle in two-dimensional space. Harmonic
multi-particle systems in Sections 5 and 6, as well as multi-particle
systems with Lennard-Jones inter-particle forces in Sections 7 and
8, respectively, constitute the main body of our paper. A\textcolor{black}{{}
final discussion of our results, indicating several applications,
is given in }Section\textcolor{black}{{} 9.}

\section{Particle swarm models }

\subsection{Active multi-particle systems}

We consider a multi-particle system of $N$ active particles \cite{paper,swarm},
enumerated by the index $i$, with equal masses $m$ in $d$ space
dimensions coupled to the internal energy $e$. The position and momentum
vectors are $\textbf{x}_{i},\textbf{p}_{i}\in\textbf{R}^{d}$ with
$i=1,\dots,N$ and the equations of motion read 
\begin{equation}
\dot{\textbf{x}}_{i}=\frac{\partial H}{\partial\textbf{p}_{i}},
\end{equation}

\begin{equation}
\dot{\textbf{p}}_{i}=-\left(1-d_{1}e\right)\frac{\partial H}{\partial\textbf{x}_{i}}-\left(\gamma-d_{2}e\right)m\frac{\partial H}{\partial\textbf{p}_{i}},
\end{equation}

\begin{equation}
\dot{e}=1-c_{1}e-c_{2}k\, e-c_{3}u\, e,
\end{equation}
where the total Hamiltonian $H=N\, h$ is the sum of kinetic $K=Nk$
and potential energy $U=Nu$ given by 
\begin{equation}
H=K+U,\qquad K=\sum_{i=1}^{N}\frac{\textbf{p}_{i}^{2}}{2m},\qquad U=\sum_{i=1}^{N}U_{i}^{(ext)}+\frac{1}{2}\sum_{i,j=1}^{N}U_{ij}^{(int)}.
\end{equation}
The potential energy $U$ of the swarm is composed of the external
potentials $U_{i}^{(ext)}$, modeling the environment of the swarm,
and of the potentials $U_{ij}^{(int)}$, describing the pairwise interactions
among the particles. The swarm model is specified by the potential
$U$ and a set of parameters $c_{1},c_{2},c_{3},d_{1},d_{2}$ and
$\gamma.$ $ $An active swarm model where $c_{2}=c_{3}$ is called
canonical. \\
\\
We note that in the fast feedback limit of internal energy a related
active swarm model was studied previously \cite{schweitzer ebeling tilch swarm}.

\subsection{Swarm dynamics for a harmonic multi-particle system }

Here we study the case where all particles are attracted with harmonic
forces, for simplicity no external forces are considered. The total
Hamiltonian then reads 
\begin{equation}
H=\sum_{i=1}^{N}\frac{\textbf{p}_{i}^{2}}{2m}+\sum_{i,j=1}^{N}\frac{m\,\omega_{0}^{2}}{4}\left(\textbf{x}_{i}-\textbf{x}_{j}\right)^{2}.\label{eq:hamiltonian_harmonic}
\end{equation}
In the center of mass frame the swarm dynamics is given by 
\begin{eqnarray}
\dot{\textbf{x}}_{i} & = & \frac{\textbf{p}_{i}}{m},\label{eq:harmonicx}\\
\dot{\textbf{p}}_{i} & = & -\left(1-d_{1}e\right)N\, m\,\omega_{0}^{2}\,\textbf{x}_{i}-\left(\gamma-d_{2}e\right)\textbf{p}_{i},\label{eq:harmonicp}\\
\dot{e} & = & 1-c_{1}e-c_{2}k\, e-c_{3}u\, e.
\end{eqnarray}
The above system of coupled nonlinear differential equations is not
easily accessible with direct analytic procedures. Nonetheless, predictions
for the system's long time behavior can be made by transforming to
macroscopic swarm variables 
\begin{equation}
e,\quad K=\sum_{i=1}^{N}\frac{\textbf{p}_{i}^{2}}{2m},\quad U=N\, m\,\omega_{0}^{2}\sum_{i=1}^{N}\frac{\textbf{x}_{i}^{2}}{2},\quad S=\frac{\sqrt{N}\omega_{0}}{2}\sum_{i=1}^{N}\textbf{x}_{i}\cdot\textbf{p}_{i}.\label{eq:swarmvariables}
\end{equation}
$K$ represents the total kinetic energy of the swarm, $U$ the total
internal potential energy, $S=Ns$ denotes the virial. We also introduce
the corresponding intensive quantities $k,u,s$ and recall $H=Nh=N(k+u)$.
These definitions of the macroscopic swarm variables are valid in
any spatial dimension $d$ of the system. The differential equations
now read

\begin{eqnarray}
{\color{black}\dot{k}} & {\color{black}=} & {\color{black}-\left(1-d_{1}\,{\color{red}{\color{black}e}}\right)2\sqrt{N}\omega_{0}s-2\left(\gamma-d_{2}\,{\color{red}{\color{black}e}}\right)\, k,}\label{swarm1}\\
{\color{black}\dot{u}} & {\color{black}=} & 2\sqrt{N}\omega_{0}{\color{black}s,}\\
{\color{black}\dot{s}} & {\color{black}=} & {\color{black}\sqrt{N}\omega_{0}k-\left(1-d_{1}\,{\color{red}{\color{black}e}}\right)\sqrt{N}\omega_{0}u-\left(\gamma-d_{2}\,{\color{red}{\color{black}e}}\right)\, s,}\\
{\color{black}{\color{black}\dot{{\color{red}{\color{red}{\color{black}e}}}}}} & {\color{black}{\color{black}=}} & 1-c_{1}e-c_{2}k\, e-c_{3}u\, e.\label{swarm2}
\end{eqnarray}
We have thus reduced the $(2dN+1)$-dimensional system $(6)-(8)$
of first order differential equations for coordinates, momenta and
internal energy to a $4$-dimen\-sional system of first order differential
equations $(10)-(13)$ for the macroscopic swarm variables. \\
 \\
In the long time limit an equilibrium state which corresponds to amorphous
swarming could be obtained by finding a stable fixed point $(k_{0},u_{0},s_{0},e_{0})$
with non-vanishing kinetic energy $k_{0}$$ $\textbf{.} Bifurcation
analysis provides us with conditions for the parameters $c_{1,}c_{2},c_{3},d_{1},d_{2},\gamma$
for this to take place, see \cite{paper,swarm}. It is obvious that
a system in a swarming equilibrium state effectively becomes thermostatted,
so a novel method of thermostatting appears to be indicated. \\
\\
Active swarm dynamics with its many parameters necessitates quite
an involved study of the various arising phenomena. In this paper
we therefore focus on simplified and more manageable time evolutions
of active swarms, setting $d_{1}=c_{1}=0$. \\

\begin{itemize}
\item Within the canonical formulation where $c_{2}=c_{3}$ we will demonstrate
that equilibrium states with fixed total energy may emerge in the
long time limit. It is precisely this phenomenon, which defines our
novel type of active ergostats. We will discuss several applications.\\
\\

\item Studying the related swarm dynamics with $d_{1}=c_{1}=0$ and $c_{3}=0$
we find equilibrium states with fixed kinetic energy. This defines
our novel type of active thermostats. We will explore features of
such a novel method of thermostatting, give its relation to the \textcolor{black}{Nosé-Hoover
thermostat and discuss several applications.}\\

\end{itemize}
We close by reminding that also static long time limits of a swarm
exist, where all particles are collapsing to a single point or freezing
according to a certain pattern. This is of no concern in the present
investigation of ergostats and thermostats, however.

\section{Ergostats and thermostats}

\subsection{Nosé-Hoover thermostat}

The novel type of ergostats and thermostats we are going to present
reminds in some aspects of the Nosé\textendash{}Hoover thermostat,
which we will review now. In order to model a system of $N$ particles
coupled to a thermal reservoir at temperature $T$ Nosé \cite{nose}
defined an extended Hamiltonian with additional canonically conjugated
degrees of freedom $s,P_{s}$ representing the heat bath; also a parameter
$Q$ named Nosé mass parameter was introduced.\textcolor{black}{{} The
equations of motion in the so called }Nosé - Hoover form \cite{hoover}
\textcolor{black}{are given by }

\begin{eqnarray}
\dot{\textbf{x}}_{i} & = & \frac{\textbf{p}_{i}}{m},\\
\dot{\textbf{p}_{i}} & = & -\frac{\partial U}{\partial\mathbf{x_{i}}}-\xi\,\mathbf{p_{i}},\\
\dot{\xi} & = & \frac{1}{\tau^{2}}\,\left(\frac{k}{k_{0}}-1\right).
\end{eqnarray}
Here $\xi=\frac{P_{s}}{Q}$ acts like an extended system feedback
force controlling the kinetic energy $k$. The Temperature $T$ of
the system is related to the average kinetic energy $k_{0}$ by $T=\frac{2}{d}\, k_{0}$,
the relaxation time $\tau$ is defined by $\tau^{2}=\frac{Q}{2Nk_{0}}$.
\\
 \\
Nosé proved analytically that the microcanonical probability measure
on the extended variable phase space reduces to a canonical probability
measure on the physical variable phase space $(\mathbf{x}_{i},\mathbf{p_{i}})$.
The Nosé\textendash{}Hoover thermostat has been commonly used as one
of the most accurate and efficient methods for constant-temperature
molecular dynamics simulations.

\subsection{Active ergostat and ergostatting in the fixed point }

Substituting into the active multi-particle system $(1)-(3)$ the
special parameter values 
\begin{equation}
c_{1}=0,\quad c_{2}=c_{3}=\frac{1}{h_{0}\tau_{2}},\quad d_{1}=0,\quad d_{2}=\frac{1}{\left(\tau_{1}\right)^{2}},\quad\gamma=\frac{\tau_{2}}{\left(\tau_{1}\right)^{2}},
\end{equation}
as well as transforming variables 
\begin{equation}
e\rightarrow\xi=\gamma-d_{2}\, e\label{eq:rescaling}
\end{equation}
we arrive \textcolor{black}{at equations of motion which are of a
generalized }Nosé - Hoover form

\begin{eqnarray}
\dot{\textbf{x}}_{i} & = & \frac{\textbf{p}_{i}}{m},\label{eq:cat1}\\
\dot{\textbf{p}_{i}} & = & -\frac{\partial U}{\partial\mathbf{x_{i}}}-\xi\,\mathbf{p_{i}},\label{cat2}\\
\dot{\xi} & = & \frac{1}{\left(\tau_{1}\right)^{2}}\left(\frac{h}{h_{0}}-1\right)-\frac{h}{\tau_{2}\, h_{0}}\,\xi.\label{cat3}
\end{eqnarray}
\textcolor{black}{The} evolution equations may have an attractive
fixed point in which the averaged total energy per particle $h=h_{0}$
becomes sharply fixed. This specific form of ergostat we would like
to call the\textcolor{black}{{} active ergostat ($AE$). }\\
\\
It should be remarked that for simplicity we defined the $\dot{\xi}$
equation (\ref{cat3}) under the assumption that $h_{0},\, h>0\text{}$.
For negative values of $h_{0}$ and possibly also for negative $h$
appropriate sign flips have to be added, see the discussion in section
3.2.\\

\subsection{Active thermostat and thermostatting in the fixed point }

At this place we mention an interesting variant of the above derivation,
which lead to (\ref{cat3}). For the swarm evolution $(1)-(3)$ we
consider now $c_{1}=c_{3}=d_{1}=0$ and have 
\begin{equation}
\dot{\textbf{x}}_{i}=\frac{\partial H}{\partial\textbf{p}_{i}},
\end{equation}

\begin{equation}
\dot{\textbf{p}}_{i}=-\frac{\partial H}{\partial\textbf{x}_{i}}-\left(\gamma-d_{2}e\right)m\frac{\partial H}{\partial\textbf{p}_{i}},
\end{equation}

\begin{equation}
\dot{e}=1-c_{2}k\, e.\label{eq:edot}
\end{equation}
\textcolor{black}{Different from} $(3)$\textcolor{black}{, where
the internal energy couples to the total energy, here in (\ref{eq:edot})
the internal energy couples to the kinetic energy only. This kind
of dynamics was introduced and explored in several applications by
}\cite{schw und ebe,schweitzer book}.\textcolor{black}{{}} Choosing
similar parameter values as before 
\begin{equation}
c_{2}=\frac{1}{k_{0}\tau_{2}},\quad d_{2}=\frac{1}{\left(\tau_{1}\right)^{2}},\quad\gamma=\frac{\tau_{2}}{\left(\tau_{1}\right)^{2}}
\end{equation}
we again transform variables according to (\ref{eq:rescaling}) and
finally are arriving \textcolor{black}{at }

\begin{eqnarray}
\dot{\textbf{x}}_{i} & = & \frac{\textbf{p}_{i}}{m},\label{anh1}\\
\dot{\textbf{p}_{i}} & = & -\frac{\partial U}{\partial\mathbf{x_{i}}}-\xi\,\mathbf{p_{i}},\label{anh2}\\
\dot{\xi} & = & \frac{1}{\left(\tau_{1}\right)^{2}}\left(\frac{k}{k_{0}}-1\right)-\frac{k}{\tau_{2}\, k_{0}}\,\xi.\label{anh3}
\end{eqnarray}
\\
We identify the extended system feedback force $\xi$ of the Nosé
- Hoover thermostat with the ``internal energy'' variable $e$ of
active Brownian motion (apart of the rescaling by $-d_{2}$ and shifting
by $\gamma$, see (\ref{eq:rescaling})). It is worth mentioning the
books \cite{ebeling book,klages book} where a related relationship
has been addressed as well.\\
 \\
\textcolor{black}{In contrast to the usual Nosé-Hoover case} the \textcolor{black}{active}
\textcolor{black}{thermostat} evolution may have an attractive fixed
point in which the averaged kinetic energy per particle $k=k_{0}$
becomes sharply fixed which allows us to define a temperature $T$.
This specific form of thermostat we would like to call the \textcolor{black}{active
thermostat ($AT$). }\\
\\
It is well known that for small systems the dynamics of the Nosé -
Hoover thermostat is nonergodic \cite{hoover,posch chaos} and trajectory
averages do not generally agree with the corresponding phase space
averages.  The question of ergodicity can also be addressed in our
present work. We remark, however, that we primarily are interested
in large particle systems, see the main body of our paper and sections
5-8. We therefore share viewpoints of Khinchin on the key role of
the many degrees of freedom and the (almost) complete irrelevance
of ergodicity \cite{khinchin,vulpiani book}. Indeed, snapshots of
the histogram of the momentum distribution for an active thermostat
and a harmonic multi-particle system, Fig. 5, show nice agreement
with a Gaussian shape, formally expected in the infinite system limit.
Concerning the active thermostat of a single harmonic oscillator,
section 4, already in previous work on active particles \cite{swarm,paper}
several bifurcation phenomena were studied and limit cycles were found
appearing after a Hopf-bifurcation point. Further studies could elaborate
on this and be subject of a similar analysis as in \cite{posch chaos,sprott},
where a highly complicated multi-part phase-space structure was seen.\\
\\
In the remainder of this paper we present analytic as well as numerical
studies to explain and demonstrate features of \textcolor{black}{the
active} \textcolor{black}{ergostat and active thermostat for small
as well as large systems} with either harmonic or Lennard-Jones forces,
respectively.\\

\section{Active ergostat and active thermostat for a single particle}

At the beginning we study the two-dimensional motion of a \textbf{single}
active particle in a harmonic and Lennard--Jones potential, respectively.
We focus on stationary motion - implying constant velocity - and investigate
possible circular orbits and their stability. It is convenient to
use polar coordinates $r$ and $\phi$ with corresponding unit vectors
$\uv_{r}$ and $\uv_{\phi}$, time-derivatives are $\dotuv_{r}=\dot{\phi}\,\uv_{\phi}$
and $\dotuv_{\phi}=-\dot{\phi}\,\uv_{r}$. For the position and momentum
we have\textbf{ 
\begin{equation}
{\bf x}=r\,\uv_{r},\qquad{\bf p}=p_{r}\,\uv_{r}+p_{\phi}\,\uv_{\phi}.
\end{equation}
}We cast the equations of motions 
\begin{equation}
\dot{{\bf x}}=\frac{1}{m}{\bf p}=\frac{1}{m}(p_{r}\,\uv_{r}+p_{\phi}\,\uv_{\phi})
\end{equation}
\begin{equation}
\dot{{\bf {\bf p}}}=-\frac{dU}{dr}\,\uv_{r}-\xi{\bf p}=\left(-\frac{dU}{dr}-\xi p_{r}\right)\uv_{r}-\xi p_{\phi}\,\uv_{\phi}
\end{equation}
into their corresponding form in polar coordinates and get (after
eliminating $\dot{\phi}$) 
\begin{equation}
m\dot{r}=p_{r},\qquad\dot{p}_{r}=\frac{p_{\phi}^{2}}{mr}-\frac{dU}{dr}-\xi p_{r},\qquad\dot{p}_{\phi}=\frac{p_{r}p_{\phi}}{mr}-\xi p_{\phi}.\label{eq:equmotion1}
\end{equation}
Finally the time evolution of $\xi$ is added, which for the $AE$
case reads 
\begin{equation}
\dot{\xi}=\frac{1}{\left(\tau_{1}\right)^{2}}\left(\frac{h}{h_{0}}-1\right)-\frac{h}{\tau_{2}\, h_{0}}\xi,\quad h=\frac{p_{r}^{2}+p_{\phi}^{2}}{2m}+U,\label{equmotioncatxi}
\end{equation}
while for the $AT$ case it is given by 
\begin{equation}
\dot{\xi}=\frac{1}{\left(\tau_{1}\right)^{2}}\left(\frac{k}{k_{0}}-1\right)-\frac{k}{\tau_{2}\, k_{0}}\xi,\quad k=\frac{p_{r}^{2}+p_{\phi}^{2}}{2m}.\label{equmotionanhxi}
\end{equation}
$ $\\
In order to reach stationarity and circular motion we are looking
for stationary points $p\rightarrow p_{r0},\; p_{\phi}\rightarrow p_{\phi0},\;\xi\rightarrow\xi_{0}$.
We demand 
\begin{equation}
p_{r0}=0,\qquad\xi_{0}=0\label{stat}
\end{equation}
and given the potential $U(r)$ are searching for solutions $r_{0}$
and $p_{\phi0}$ which for the $AE$ case fulfill 
\begin{equation}
p_{\phi0}=\sqrt{mr_{0}U'[r_{0}]},\qquad h_{0}=U[r_{0}]+p_{\phi0}^{2}/2m\quad(AE),\label{cat}
\end{equation}
while in the $AT$ case 
\begin{equation}
p_{\phi0}=\sqrt{mr_{0}U'[r_{0}]},\qquad k_{0}=p_{\phi0}^{2}/2m\quad(AT).\label{anh}
\end{equation}
We linearize the equations of motion around the stationary points
and discuss stability. Without solving the linear dynamical system
directly we use the Routh\textendash{}Hurwitz test as an efficient
recursive algorithm to determine whether all the roots of the characteristic
polynomial have negative real parts. The Routh-Hurwitz stability criterion
proclaims that all first column elements of the so called Routh array
have to be of the same sign.

\subsection{Harmonic potential}

For the harmonic oscillator potential with coupling constant $\kappa$
\begin{equation}
U(r)=\frac{\kappa}{2}r^{2}
\end{equation}
in the $AE$ case for all $h_{0}>0$ there are solutions to (\ref{cat}),
guaranteeing stationarity and circular motion 
\begin{equation}
r_{0}=\sqrt{h_{0}/\kappa},\quad p_{\phi0}=\sqrt{mh_{0}}\qquad\qquad\text{(\ensuremath{AE})}.
\end{equation}
In the $AT$ case for all $k_{0}>0$ one finds corresponding solutions
to (\ref{anh}) 
\begin{equation}
r_{0}=\sqrt{2k_{0}/\kappa},\quad p_{\phi0}=\sqrt{2mk_{0}}\qquad\qquad\text{(\ensuremath{AT})}.
\end{equation}
The Routh-Hurwitz stability criterion predicts (marginal) stability
$(+,+,+,0,+)$ for the $AE$ and full stability $(+,+,+,+,+)$ for
the $AT$ case.

\subsection{Lennard--Jones potential}

The Lennard--Jones potential has the form 
\begin{equation}
U(r)=\epsilon\left(\left(\frac{a}{r}\right)^{12}-2\left(\frac{a}{r}\right)^{6}\right),
\end{equation}
where $a$ is the distance at which the potential $U$ reaches its
minimal value $-\epsilon=U(a)$. Stationary and circular motion arises
in the $AE$ case (\ref{cat}) with 
\begin{alignat}{1}
r_{0} & =a\left(\frac{2}{5}\left(1+\sqrt{1-{5h_{0}}/{4\epsilon}}\right)\right)^{-1/6},\label{eq:cat2r}\\
p_{\phi0} & =\frac{2}{5}\sqrt{3m\left(5h_{0}+2\epsilon\left(1+\sqrt{1-{5h_{0}}/{4\epsilon}}\right)\right)}\label{eq:cat2phi}
\end{alignat}
for two different regimes of the total energy. One solution exists
for positive $h_{0}$, where $0<h_{0}\le\frac{4}{5}\epsilon$, and
the Routh-Hurwitz analysis predicts (marginally) stability $(+,+,+,0,+).$
\\
\\
The second (marginally) stable solution arises for negative $h_{0}$,
where $-\epsilon<h_{0}<0$. It should be noted, however, that in order
to reach such a fixed point for negative $h_{0}$ (for simplicity
we are also assuming $h(t)<0$) the sign of the first term on the
right hand side of the $\dot{\xi}$ $ $equation (\ref{equmotioncatxi})
has to be flipped 
\begin{equation}
\dot{\xi}=-\frac{1}{\left(\tau_{1}\right)^{2}}\left(\frac{h}{h_{0}}-1\right)-\frac{h}{\tau_{2}\, h_{0}}\xi,\qquad h_{0},h<0
\end{equation}
Indeed, in this case again the Routh-Hurwitz analysis predicts marginal
stability $(+,+,+,0,+)$.\\
\\
Next we turn to the $AT$ case case, where for $0<k_{0}\le\frac{2}{3}\epsilon$
a solution exists with 
\begin{equation}
r_{0}=a\left(\frac{1}{2}\left(1+\sqrt{1-2k_{0}/3\epsilon}\right)\right)^{-1/6},\qquad p_{\phi0}=\sqrt{2k_{0}m},
\end{equation}
for which the Routh-Hurwitz analysis predicts stability $(+,+,+,+,+).$
\\
\\
Further unstable solutions in the $AE$ and $AT$ case exist, but
for simplicity will not be discussed here.

\section{Active ergostat for a harmonic multi-particle system}

\begin{figure}
\begin{center}
\includegraphics[width=10cm]{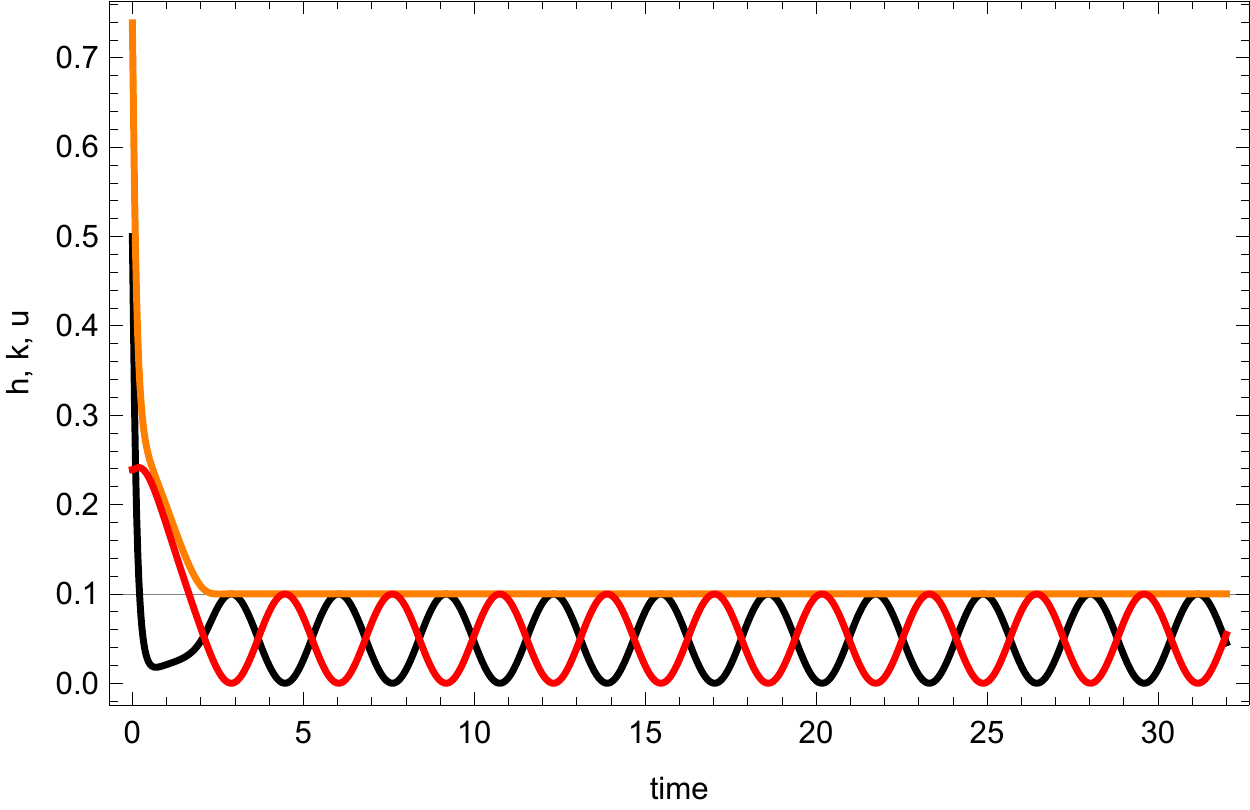}
\end{center}
\caption{\label{fig:energiesfigs_h_CAT512.pdf}Plot of total energy $h$ (orange),
kinetic energy $k$ (black) and potential energy $u$ (red) for an
active ergostat and a harmonic multi-particle system. Simulation parameters
are: $N=512,d=2,m=1,\omega_{0}=1,\tau_{1}=0.1,\tau_{2}=0.05,T_{init}=0.5,h_{0}=0.1$}
\end{figure}

In this section we perform the detailed numerical simulation of an
active ergostat for a \textcolor{black}{N-particle system }with harmonic
forces described by the Hamiltonian $H=\sum\frac{\textbf{p}_{i}^{2}}{2m}+\sum_{i,j=1}^{N}\frac{m\,\omega_{0}^{2}}{4}\left(\textbf{x}_{i}-\textbf{x}_{j}\right)^{2}$.
In the active ergostat case (\ref{eq:harmonicx}-\ref{eq:harmonicp})
and (\ref{cat3}) we have 
\begin{equation}
\dot{\textbf{x}}_{i}=\frac{\textbf{p}_{i}}{m},
\end{equation}
\begin{equation}
\dot{\textbf{p}}_{i}=-N\, m\,\omega_{0}^{2}\,\textbf{x}_{i}-\xi\textbf{p}_{i},
\end{equation}
\begin{equation}
\dot{\xi}=\frac{1}{\left(\tau_{1}\right)^{2}}\left(\frac{h}{h_{0}}-1\right)-\frac{h}{\tau_{2}\, h_{0}}\,\xi
\end{equation}
We study a system with $N=512$ particles. Initial conditions are
prepared in such a way that CMS coordinates and momenta are vanishing,
the coordinates are chosen randomly from within a circle of fixed
length. The momenta are taken randomly from a Maxwellian distribution,
according to some chosen initial temperature $T_{init}$. \\
\\
The system quickly relaxes to a fixed point with constant total energy
$h_{0}.$ The extended system feedback force $\xi$ decreases rapidly
without oscillations and is vanishing in good approximation. In contrast
to it the kinetic energy $k$ and the internal potential energy $u$
are oscillating permanently. Their sum, however, is stabilized at
the chosen fixed point value $h_{0}$, as can be seen in Fig. \ref{fig:energiesfigs_h_CAT512.pdf}
\\
\\
See Supplemental Material at \cite{download} for videos of the swarm evolution
together with the corresponding histograms of the momenta.
The swarm is seen to be oscillating regularly.
Each time after a phase of expansion,
for a short moment, all particles come to complete rest. Subsequently
the swarm continues contracting towards the origin, where it starts
expanding again.

\section{Active thermostat for a harmonic multi-particle system}

\begin{figure}
\begin{center}
\includegraphics[width=10cm]{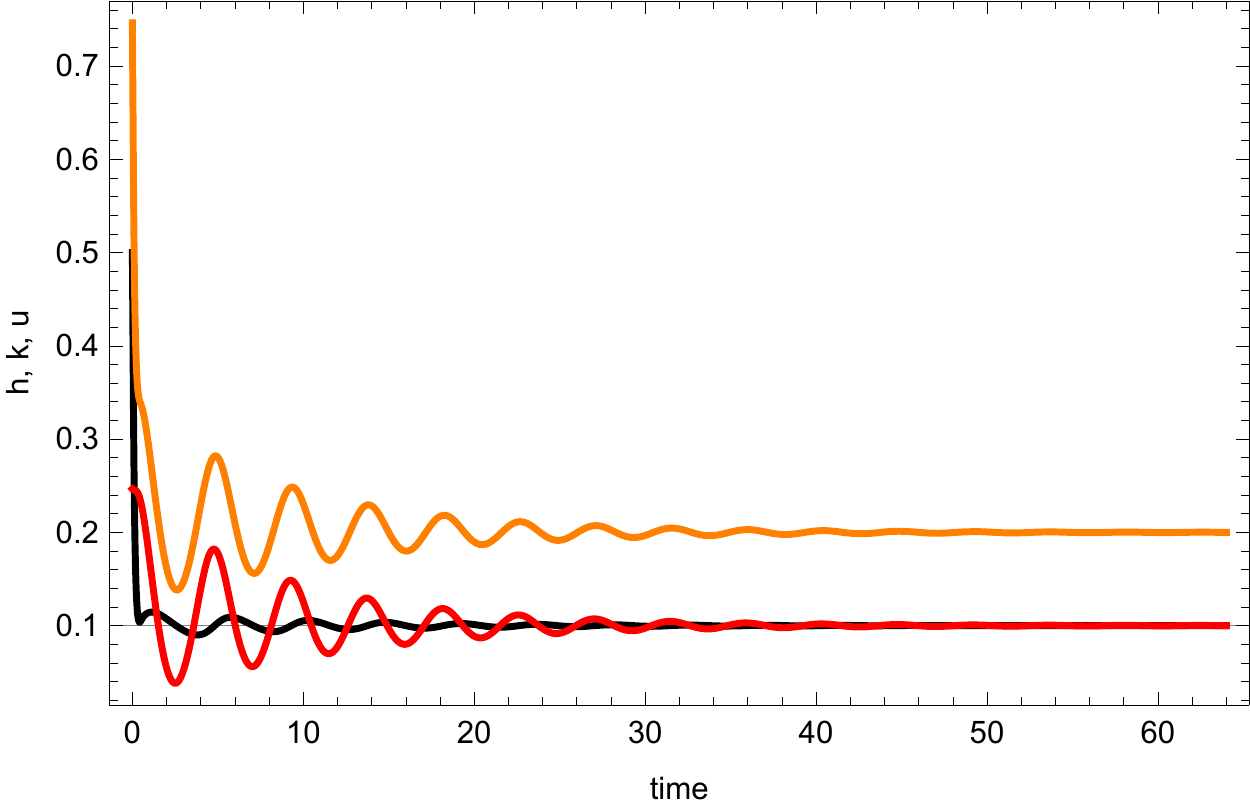}
\end{center}
\caption{\label{fig:energiesfigs_h_ANH512.pdf}Plot of total energy $h$ (orange),
kinetic energy $k$ (black) and potential energy $u$ (red) for an
active thermostat and a harmonic multi-particle system. Simulation
parameters are: $N=512,d=2,m=1,\omega_{0}=1,\tau_{1}=0.1,\tau_{2}=0.05,T_{init}=0.5,T_{final}=0.1$}
\end{figure}

In the active thermostat case the time evolution of $\xi$ is given
by (\ref{anh3})

\begin{equation}
\dot{\xi}=\frac{1}{\left(\tau_{1}\right)^{2}}\left(\frac{k}{k_{0}}-1\right)-\frac{k}{\tau_{2}\, k_{0}}\xi,\quad k=\frac{1}{N}\sum_{i=1}^{N}\frac{\textbf{p}_{i}^{2}}{2m}
\end{equation}
In Fig \ref{fig:energiesfigs_h_ANH512.pdf} the kinetic (black), potential
(red) and total energies (orange) are plotted, all are oscillating
and are exponentially damped.

\begin{figure}
\begin{center}
\includegraphics[width=10cm]{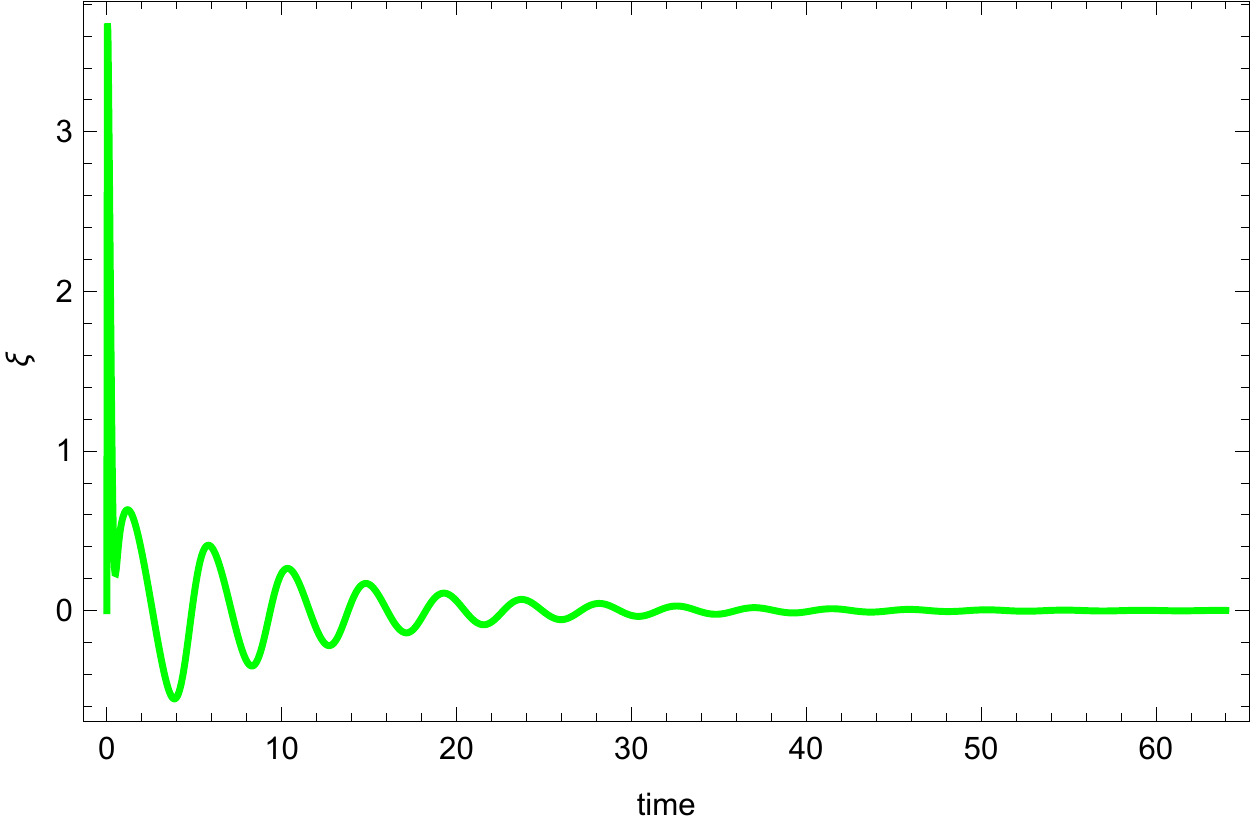}
\end{center}
\caption{\label{fig:xifig_h_ANH512.pdf}The extended system feedback force
$\xi$ for an active thermostat and a harmonic multi-particle system.
Simulation parameters are: $N=512,d=2,m=1,\omega_{0}=1,\tau_{1}=0.1,\tau_{2}=0.05,T_{init}=0.5,T_{final}=0.1$}
\end{figure}

The extended system feedback force $\xi$ shows oscillatory behavior
and exponential decrease toward $0$, see Fig. \ref{fig:xifig_h_ANH512.pdf}.

\begin{figure}
\begin{center}
\includegraphics[width=10cm]{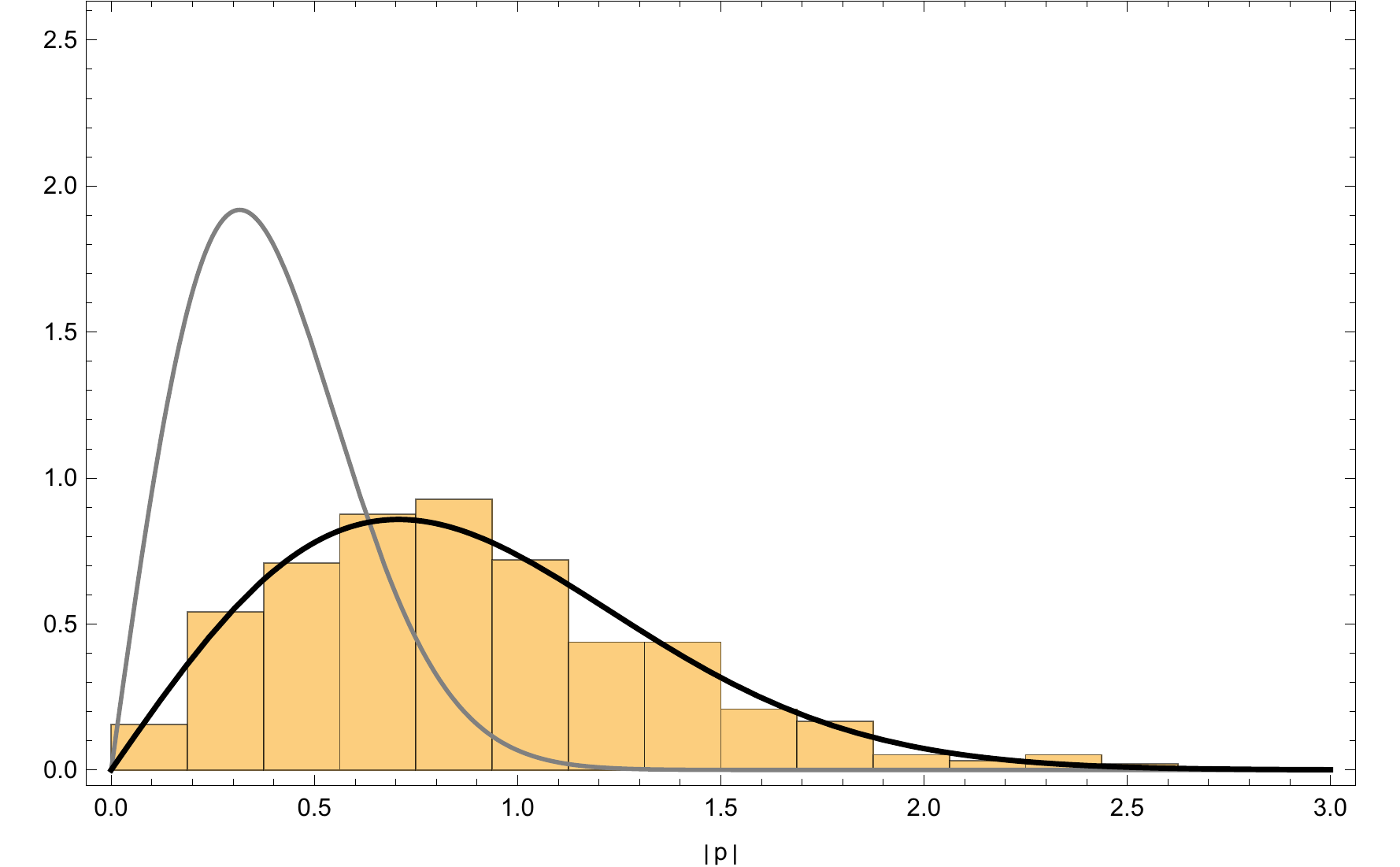}
\end{center}
\caption{\label{fig:histofirst_h_ANH512.pdf-1}Histogram of the initial momentum
distribution for an active thermostat and a harmonic multi-particle
system. Simulation parameters are: $N=512,d=2,m=1,\omega_{0}=1,\tau_{1}=0.1,\tau_{2}=0.05,T_{init}=0.5$
(corresponding to the black line), $T_{final}=0.1$ (corresponding
to the gray line)}
\end{figure}

In Fig. \ref{fig:histofirst_h_ANH512.pdf-1} the N-particle histogram
of the momentum distribution at the initial moment of the simulation
is presented. Due to our choice of initial conditions the histogram
of the momenta is following closely a pattern related to a Maxwellian
distribution (black solid line), corresponding to the initial temperature
$T_{init}$.

\begin{figure}
\begin{center}
\includegraphics[width=10cm]{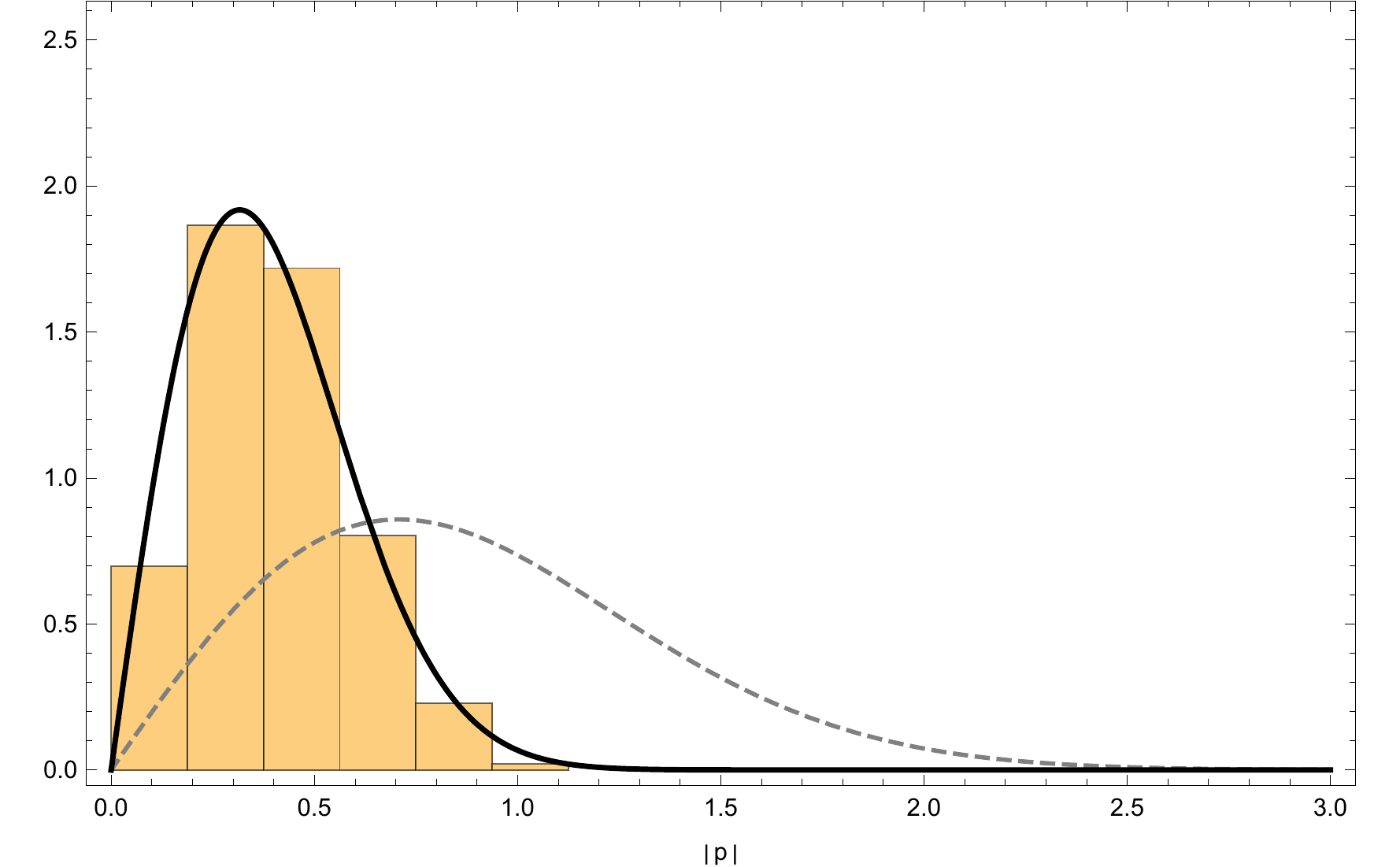}
\end{center}
\caption{\label{fig:histolast_h_ANH512.pdf}Snapshot of the histogram of the
momentum distribution for an active thermostat and a harmonic multi-particle
system after thermostatting. Simulation parameters are: $N=512,d=2,m=1,\omega_{0}=1,\tau_{1}=0.1,\tau_{2}=0.05,T_{init}=0.5$
(corresponding to the gray dashed line), $T_{final}=0.1$ (corresponding
to the black line)}
\end{figure}

Only a few time steps after the start of the simulation the system
reaches the final temperature $T_{final}$, we give a snapshot of
the histogram in Fig. \ref{fig:histolast_h_ANH512.pdf}. The histogram
shows a pattern related to a Maxwellian distribution (black solid
line), which is corresponding to the final temperature $T_{final}$.
See Supplemental Material at \cite{download} for videos of the swarm evolution
together with the corresponding histograms of the momenta.
\\
\\
A precise understanding of the rather complex numerical results in
the preceding and the present section can be obtained if we study
the \textcolor{black}{harmonic N-particle system i}n terms of the
macroscopic swarm variables (\ref{eq:swarmvariables} - \ref{swarm2}).
For the active ergostat case we have 

\begin{flushleft}
\begin{eqnarray}
\dot{k} & = & {\color{black}-2\sqrt{N}\omega_{0}s-2k\,\xi,}\\
\dot{u} & = & 2\sqrt{N}\omega_{0}s\\
\dot{s} & = & \sqrt{N}\omega_{0}k-\sqrt{N}\omega_{0}u-s\,\xi\\
\dot{\xi} & = & \frac{1}{\left(\tau_{1}\right)^{2}}\left(\frac{h}{h_{0}}-1\right)-\frac{h}{\tau_{2}\, h_{0}}\,\xi,\qquad h=k+u
\end{eqnarray}
Linearizing around the stationary point $k\rightarrow k_{0},u\rightarrow u_{0},s\rightarrow s_{0},\xi\rightarrow\xi_{0}$
where
\par\end{flushleft}

\begin{flushleft}
\begin{equation}
k_{0}=u_{0}=\frac{h_{0}}{2},\quad s_{0}=0,\quad\xi_{0}=0
\end{equation}
the eigenvalues of the characteristic polynomial read
\par\end{flushleft}

\begin{equation}
\left\{ -\frac{\sqrt{\text{\ensuremath{\tau_{1}}}^{2}-4\text{\ensuremath{\tau_{2}}}^{2}}+\tau_{1}}{2\text{\ensuremath{\tau_{1}}}\text{\text{\ensuremath{\tau_{2}}}}},\frac{\sqrt{\text{\ensuremath{\tau_{1}}}^{2}-4\text{\text{\ensuremath{\tau_{2}}}}^{2}}-\tau_{1}}{2\text{\ensuremath{\tau_{1}}}\text{\text{\ensuremath{\tau_{2}}}}},-2i\sqrt{N}\text{\ensuremath{\omega_{0}}},2i\sqrt{N}\text{\ensuremath{\omega_{0}}}\right\} 
\end{equation}
\\
The solutions of the linearized system can straightforwardly be obtained,
we prefer, however, to give an easy example. We choose
\begin{equation}
\tau_{1}=1,\,\tau_{2}=\frac{1}{2},\,\omega_{0}=\frac{1}{\sqrt{N}}\label{eq:paprameter}
\end{equation}
so that the eigenvalues are simply $\{-1,-1,+i,-i).$ With initial
conditions $(k(0),u(0),s(0),\xi(0))=(1,0,0,0)$ and $h_{0}=1$ we
find

\begin{eqnarray}
k & = & \frac{1}{2}+\frac{1}{25}e^{-t}\left(15t+11\right)-\frac{2}{25}\left(\sin(2t)-7\cos(2t)\right)\\
u & = & \frac{1}{2}+\frac{2}{25}e^{-t}\left(5t+7\right)+\frac{2}{25}\left(\sin(2t)-7\cos(2t)\right)\\
s & = & -\frac{1}{25}e^{-t}\left(5t+2\right)+\frac{2}{25}\left(7\sin(2t)+\cos(2t)\right)\\
\xi & = & e^{-t}t
\end{eqnarray}
We see that in the active ergostat case $k$ and $ $$u$ have exponentially
in the time decreasing contributions but also undamped oscillations.
For the total energy $h=k+u$ the undamped oscillatory parts cancel
out. Conversely $\xi$ is just exponentially decreasing without oscillations.\\
~\\
For the active thermostat case the \textcolor{black}{evolution equations
are given by}

\begin{eqnarray}
\dot{k} & = & {\color{black}-2\sqrt{N}\omega_{0}s-2k\,\xi,}\\
\dot{u} & = & 2\sqrt{N}\omega_{0}s\\
\dot{s} & = & \sqrt{N}\omega_{0}k-\sqrt{N}\omega_{0}u-s\,\xi\\
\dot{\xi} & = & \frac{1}{\left(\tau_{1}\right)^{2}}\left(\frac{k}{k_{0}}-1\right)-\frac{k}{\tau_{2}\, k_{0}}\xi
\end{eqnarray}
Linearizing around the stationary point $k\rightarrow k_{0},u\rightarrow u_{0},s\rightarrow s_{0},\xi\rightarrow\xi_{0}$
where

\begin{equation}
k_{0}=u_{0}\quad s_{0}=0,\quad\xi_{0}=0
\end{equation}
the system again can straightforwardly be solved, yet the solutions
are of quite lengthy form. We choose again the special parameter values
(\ref{eq:paprameter}). In this case the eigenvalues become twofold
degenerate $\lambda=-1\pm i\sqrt{3}$. Considering once more the initial
conditions $(k(0),u(0),s(0),\xi(0))=(1,0,0,0)$ and choosing $k_{0}=1$
the solutions of the linearized dynamics are 

\begin{eqnarray}
k & = & 1+\frac{1}{3}e^{-t}\left(\sqrt{3}(2-3t)\sin\left(\sqrt{3}t\right)-3(t-1)\cos\left(\sqrt{3}t\right)\right)\\
u & = & 1+\frac{1}{3}e^{-t}\left(\sqrt{3}(t+1)\sin\left(\sqrt{3}t\right)-3t\cos\left(\sqrt{3}t\right)\right)\\
s & = & \frac{1}{3}e^{-t}t\left(\sqrt{3}\sin\left(\sqrt{3}t\right)+3\cos\left(\sqrt{3}t\right)\right)\\
\xi & = & -\frac{4e^{-t}(t-1)\sin\left(\sqrt{3}t\right)}{\sqrt{3}}
\end{eqnarray}
All quantities show exponentially damped oscillations.\\
\\
Finally we apply our analysis the Nosé--Hoover thermostat. In this
case one has 

\begin{flushleft}
\begin{eqnarray}
\dot{k} & = & {\color{black}-2\sqrt{N}\omega_{0}s-2k\,\xi,}\\
\dot{u} & = & 2\sqrt{N}\omega_{0}s\\
\dot{s} & = & \sqrt{N}\omega_{0}k-\sqrt{N}\omega_{0}u-s\,\xi\\
\dot{\xi} & = & \frac{1}{\left(\tau_{1}\right)^{2}}\left(\frac{k}{k_{0}}-1\right)
\end{eqnarray}
It is well known that in the Nosé--Hoover case no stable fixed points
are existing. This is easily demonstrated by linearizing around the
stationary point $k\rightarrow k_{0},u\rightarrow u_{0},s\rightarrow s_{0},\xi\rightarrow\xi_{0}$
where
\par\end{flushleft}

\begin{equation}
k_{0}=u_{0}\quad s_{0}=0,\quad\xi_{0}=0
\end{equation}
One finds the strictly imaginary eigenvalues 
\begin{equation}
\left\{ \pm i\frac{\sqrt{\sqrt{4N^{2}\text{\ensuremath{\tau_{1}}}^{4}\text{\ensuremath{\omega_{0}}}^{4}+1}+2N\text{\ensuremath{\tau_{1}}}^{2}\text{\ensuremath{\omega_{0}}}^{2}+1}}{\tau_{1}},\pm i\frac{\sqrt{-\sqrt{4N^{2}\text{\ensuremath{\tau_{1}}}^{4}\text{\ensuremath{\omega_{0}}}^{4}+1}+2N\text{\ensuremath{\tau_{1}}}^{2}\text{\ensuremath{\omega_{0}}}^{2}+1}}{\tau_{1}},\right\} 
\end{equation}
so all macroscopic swarm variables are showing undamped oscillations.

\section{Active ergostat for a multi-particle system with Lennard--Jones force}

\begin{figure}
\begin{center}
\includegraphics[width=10cm]{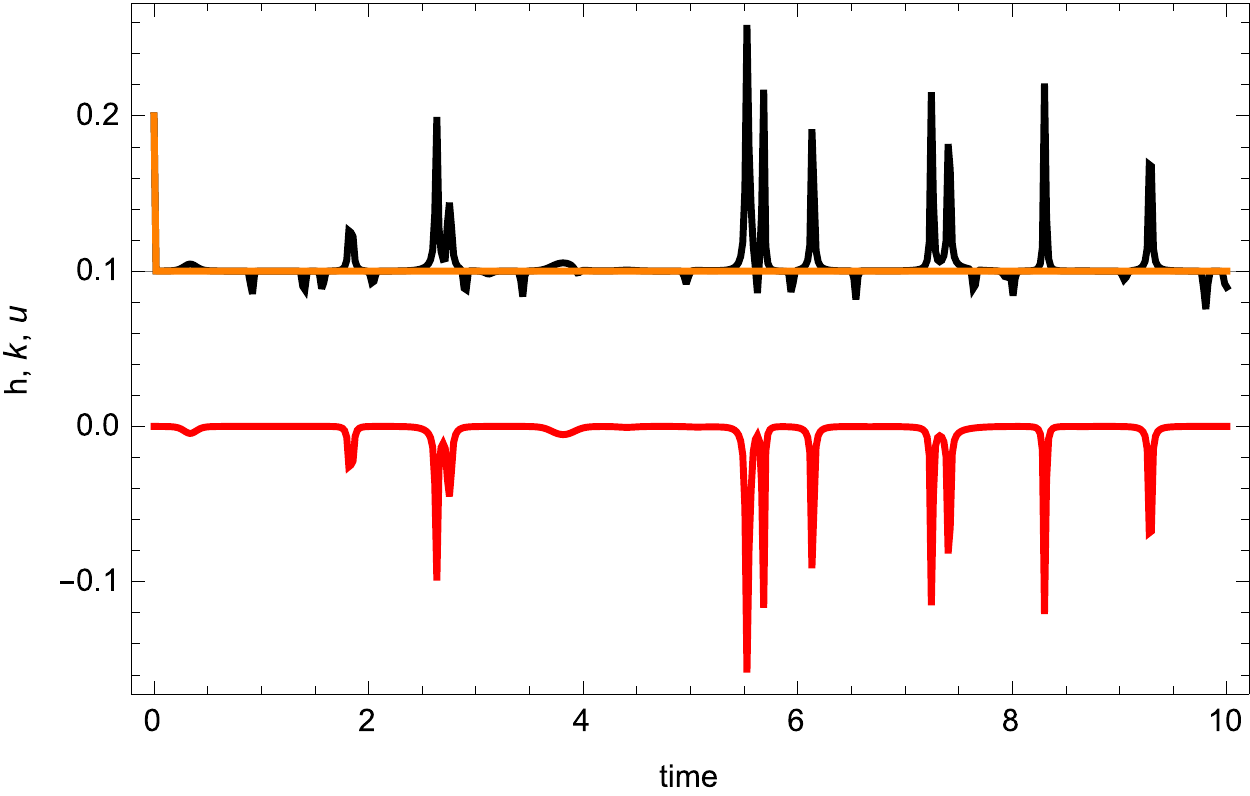}
\end{center}

\caption{\label{fig:hku21.eps}Plot of total energy $h$ (orange), kinetic
energy $k$ (black) and potential energy $u$ (red) for an active
ergostat and a Lennard--Jones multi-particle system. Simulation parameters
are: $N=8,d=2,m=1,a=0.05,\varepsilon=1,\tau_{1}=0.1,\tau_{2}=0.05,T_{init}=0.2,h_{0}=0.1$}
\end{figure}

In this section we perform the numerical simulation of an active ergostat
for a \textcolor{black}{N-particle system }with Lennard--Jones forces.
The total Hamiltonian reads 
\begin{equation}
H=\sum_{i=1}^{N}\frac{\textbf{p}_{i}^{2}}{2m}+\frac{1}{2}\sum_{i,j=1}^{N}\epsilon\left(\left(\frac{a}{\mid\textbf{x}_{i}-\textbf{x}_{j}\mid}\right)^{12}-2\left(\frac{a}{\mid\textbf{x}_{i}-\textbf{x}_{j}\mid}\right)^{6}\right).\label{eq:hamiltonlj}
\end{equation}
and the system evolves according to
\begin{equation}
\dot{\textbf{x}}_{i}=\frac{\textbf{p}_{i}}{m},\label{eq:xlj}
\end{equation}
\begin{equation}
\dot{\textbf{p}}_{i}=-\frac{\partial H}{\partial\mathbf{x_{i}}}-\xi\textbf{p}_{i},\label{eq:plj}
\end{equation}
\begin{equation}
\dot{\xi}=\frac{1}{\left(\tau_{1}\right)^{2}}\left(\frac{h}{h_{0}}-1\right)-\frac{h}{\tau_{2}\, h_{0}}\,\xi,\quad h=\frac{H}{N}.
\end{equation}
The initial conditions are prepared in such a way that CMS coordinates
and momenta are vanishing. For the simulation of a Lennard--Jones
system it is preferable to choose the initial coordinates randomly
from a regular grid within a circle of fixed radius. The initial momenta
are taken randomly from a Maxwellian distribution, corresponding to
some chosen initial temperature $T_{init}$. \\
\\
First we study ergostatting for a small particle number N=8. In Fig.
\ref{fig:hku21.eps} a plot of the kinetic (black), potential (red)
and total (orange) energy is given. 

\begin{figure}
\begin{center}
\includegraphics[width=10cm]{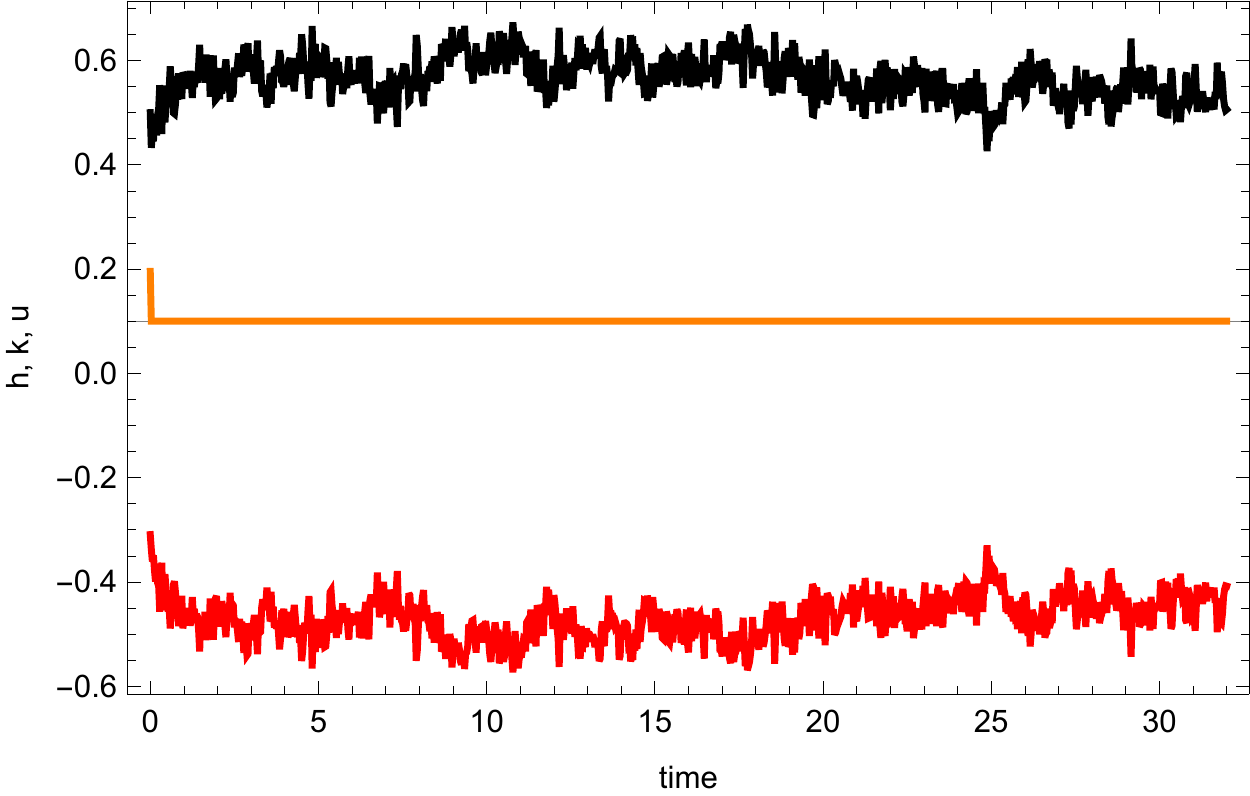}
\end{center}
\caption{\label{fig:energiesfig_LJCAT128.pdf}Plot of total energy $h$ (orange),
kinetic energy $k$ (black) and potential energy $u$ (red) for an
active ergostat and a Lennard--Jones multi-particle system. Simulation
parameters are: $N=128,d=2,m=1,a=0.05,\varepsilon=1,\tau_{1}=0.1,\tau_{2}=0.05,T_{init}=0.5,h_{0}=0.1$}
\end{figure}

We find that the total energy $h$ is quickly fixed at its required
value $h_{0}$. The clearly pronounced positive spikes in the kinetic
energy and the coinciding negative spikes in the potential energy
correspond to events where two particles find themselves sufficiently
close one to another. The potential energy of the system receives
a negative contribution which due to the ergostatting mechanism leads
to an increase of the kinetic energy, which prevents clusterization.
The small negative spikes in the kinetic energy are a consequence
of the interaction of the particles with the external potential that
is introduced to prevent the swarm from spreading apart. For simplicity
we did not include the plot of the external potential. \\
\\
The extended system feedback force $\xi$ is vanishing in good approximation
after a very short moment and the system approximately becomes Hamiltonian.
As now the conservation of the total energy is guaranteed by the Hamiltonian
dynamics itself, ergostatting due to the extended system feedback
force $\xi$ has only minor importance. \\
\\
When studying a system with $N=128$ particles the above features
and interpretations get somewhat washed out. It can clearly be seen
again that the system quickly achieves the required total energy $h_{0}.$
The kinetic energy $k$ and the averaged potential energy $u$ of
the system are fluctuating quite heavily, yet their sum $h=k+u$ is
stabilized well, see Fig. \ref{fig:energiesfig_LJCAT128.pdf}.

\begin{figure}
\begin{center}
\includegraphics[width=10cm]{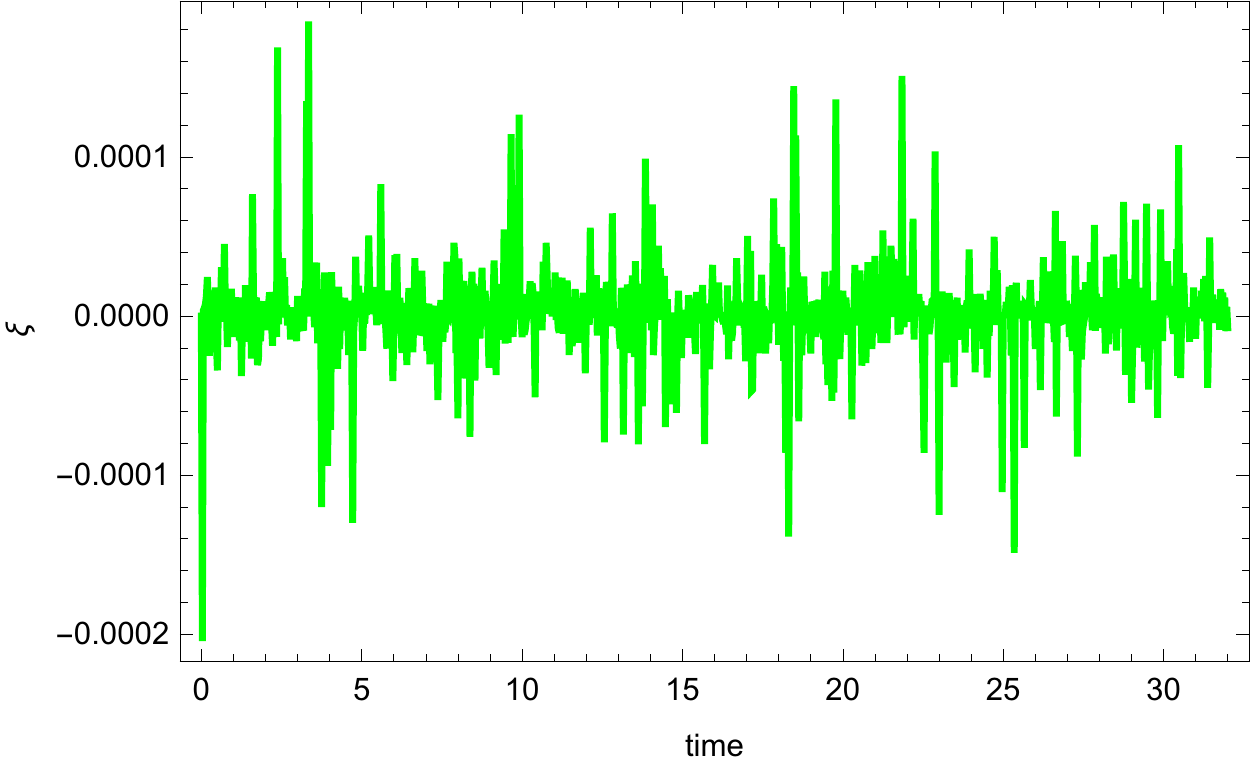}
\end{center}
\caption{\label{fig:xifig_LJCAT128.pdf}Plot of extended system feedback force
$\xi$ for an active ergostat and a Lennard--Jones multi-particle
system. Simulation parameters are: $N=128,d=2,m=1,a=0.05,\varepsilon=1,\tau_{1}=0.1,\tau_{2}=0.05,T_{init}=0.5,h_{0}=0.1$}
\end{figure}

The extended system feedback force is fluctuating at a small order
of magnitude, see Fig. \ref{fig:xifig_LJCAT128.pdf}.

See Supplemental Material at \cite{download} for videos of the swarm evolution
together with the corresponding histograms of the momenta.\\
\\

\section{Active thermostat for a multi-particle system with Lennard--Jones
force}

\textcolor{black}{We perform the simulation of an active thermostat
for a N-particle system }with the Lennard--Jones Hamiltonian (\ref{eq:hamiltonlj})
and the time evolution (\ref{eq:xlj}), (\ref{eq:plj}) and (\ref{anh3}).\\
 
\begin{figure}
\begin{center}
\includegraphics[width=10cm]{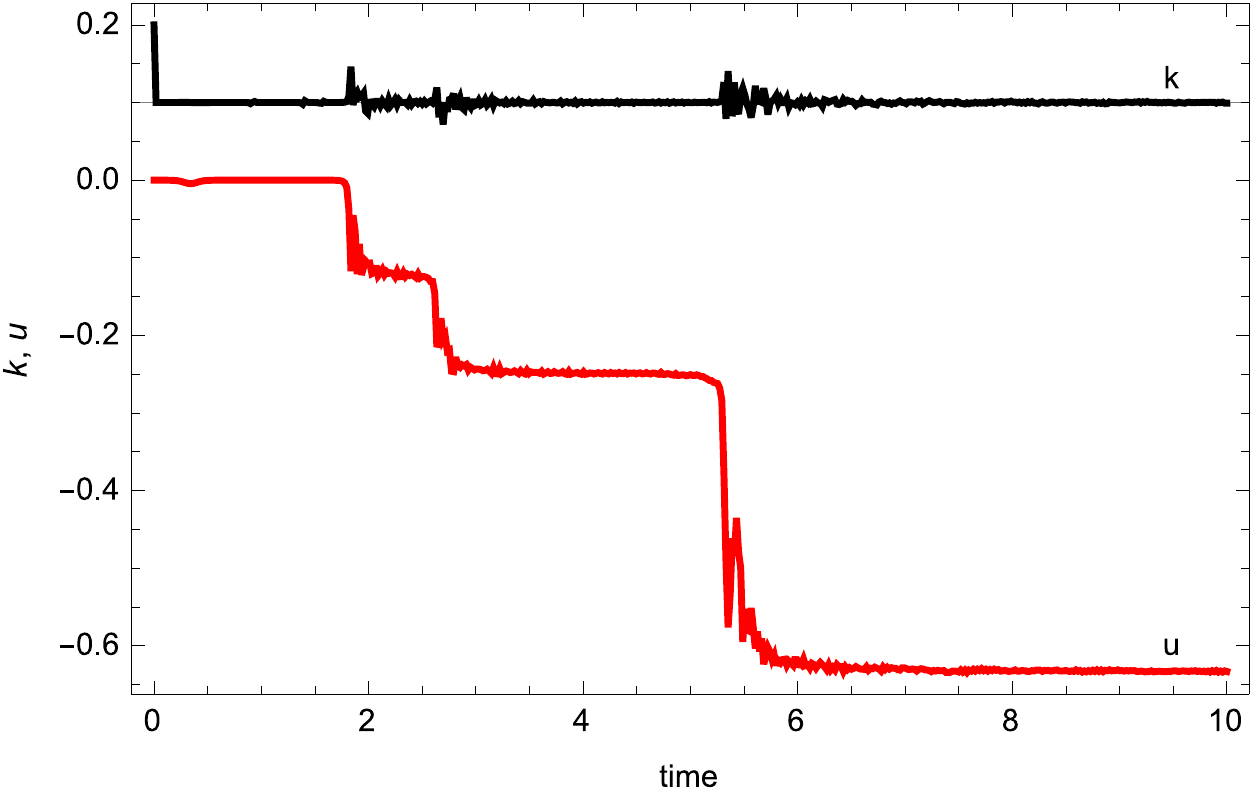}
\end{center}
\caption{\label{fig:hku20.eps}Plot of kinetic energy $k$ (black) and potential
energy $u$ (red) for an active thermostat and a Lennard--Jones multi-particle
system. Simulation parameters are: $N=8,d=2,m=1,a=0.05,\varepsilon=1,\tau_{1}=0.1,\tau_{2}=0.05,T_{init}=0.2,T_{final}=0.1$ }
\end{figure}
\begin{figure}
\begin{center}
\includegraphics[width=10cm]{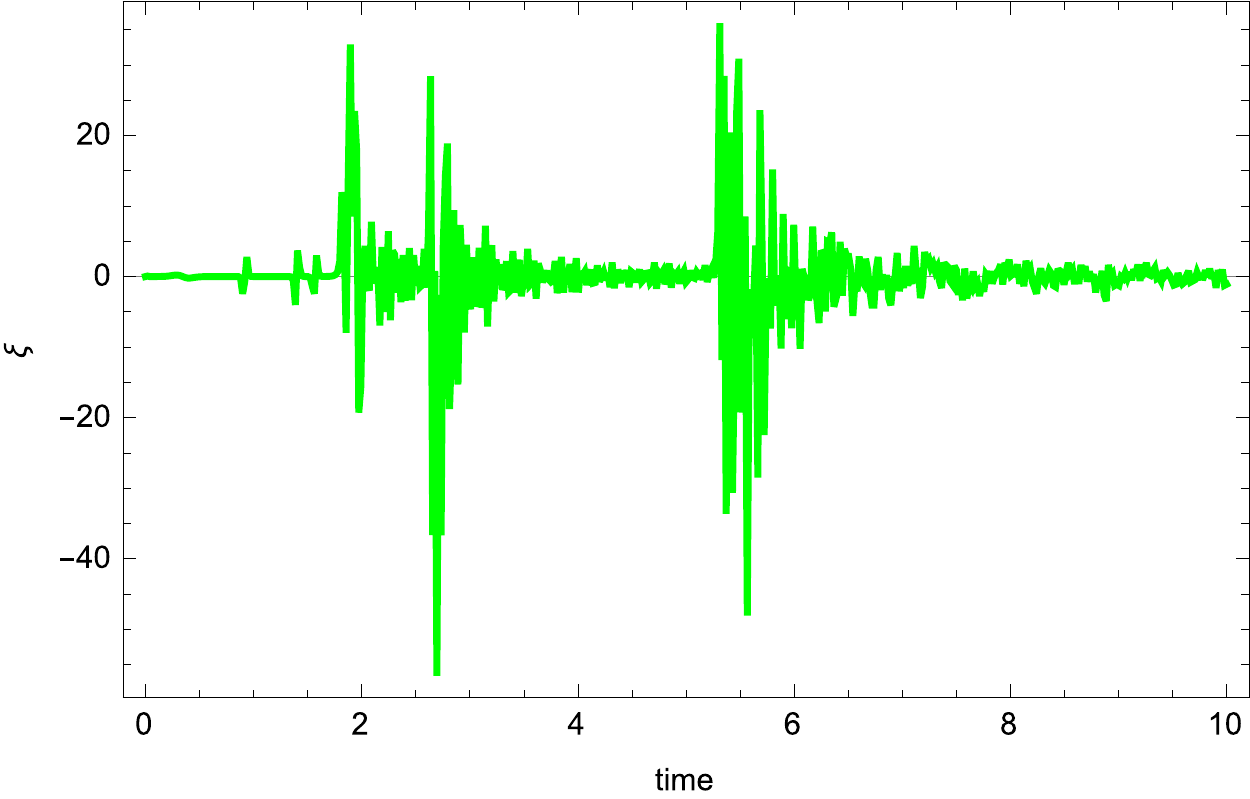}
\end{center}

\caption{\label{fig:xi0.eps}Plot of extended system feedback force $\xi$
for an active thermostat and a Lennard--Jones multi-particle system.
Simulation parameters are: $N=8,d=2,m=1,a=0.05,\varepsilon=1,\tau_{1}=0.1,\tau_{2}=0.05,T_{init}=0.2,T_{final}=0.1$ }
\end{figure}

Again we first study the case of small particle numbers, choosing
N=8. We observe that $k$ quickly reaches the prescribed stationary
value $k_{0}$, while $u$ shows synchronized stepwise transitions
towards lower values. In Fig. \ref{fig:hku20.eps} the kinetic (black)
and potential (red) energies are plotted. Each transition corresponds
to the formation of a cluster of a pair of particles or of additional
particles joining an already existing cluster. As each binding of
a particle adds an amount of negative potential energy to the system,
the total energy h decreases accordingly. If all particles would form
one big cluster, the kinetic energy of the whole system would divide
itself between the centre of mass motion / rotation of the cluster
and the vibrations of all the bound particles.\\
\\
The extended system feedback force $\xi$ is stabilizing the kinetic
energy $k$ by bursts of fluctuations, this can be seen in Fig. \ref{fig:xi0.eps}.\\

\begin{figure}
\begin{center}
\includegraphics[width=10cm]{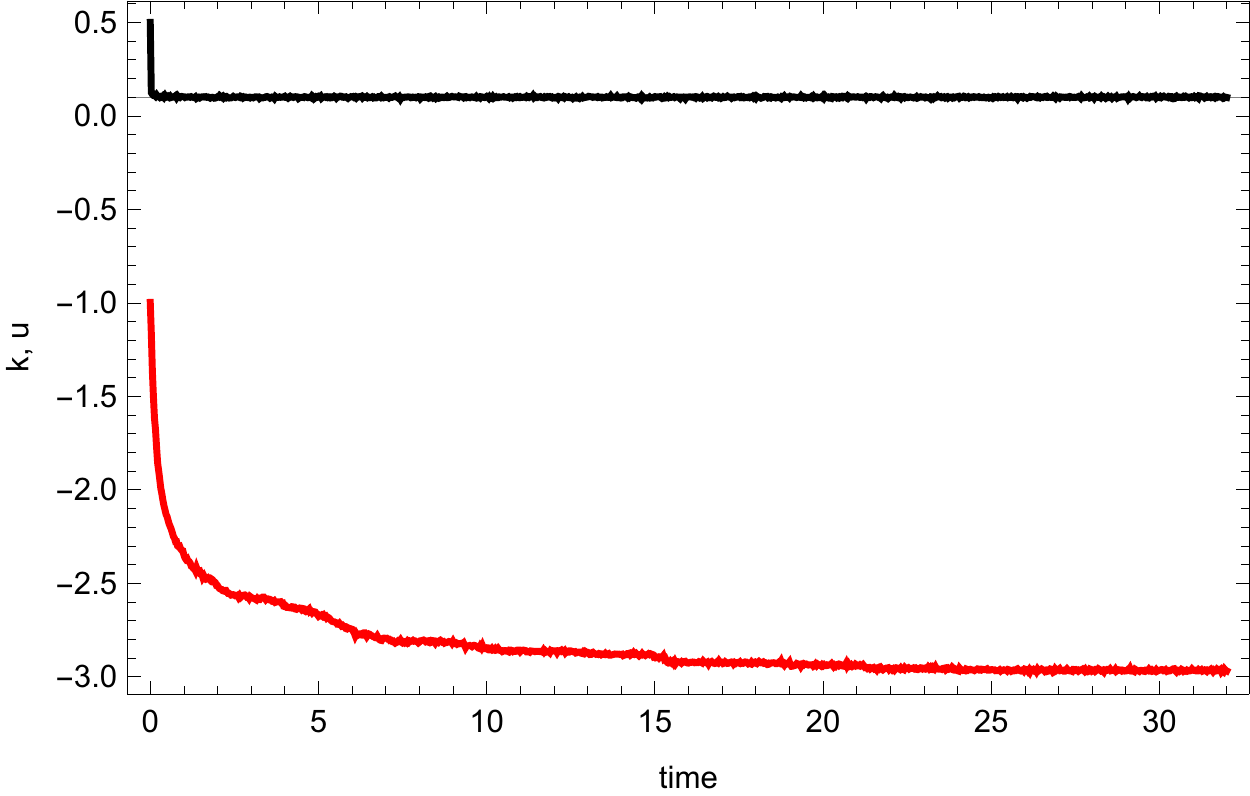}
\end{center}
\caption{\label{fig:energiesfig_lj512.pdf}Plot of kinetic energy $k$ (black)
and potential energy $u$ (red) for an active thermostat and a Lennard--Jones
multi-particle system. Simulation parameters are: $N=512,d=2,m=1,a=0.05,\varepsilon=1,\tau_{1}=0.1,\tau_{2}=0.05,T_{init}=0.5,T_{final}=0.1$}
\end{figure}

In a simulation of the Lennard-Jones gas with particle number N=512
the main features of our analysis persist, see Fig. \ref{fig:energiesfig_lj512.pdf}
for plots of the kinetic (black) and potential (red) energies. \\

\begin{figure}
\begin{center}
\includegraphics[angle=-90,width=12cm]{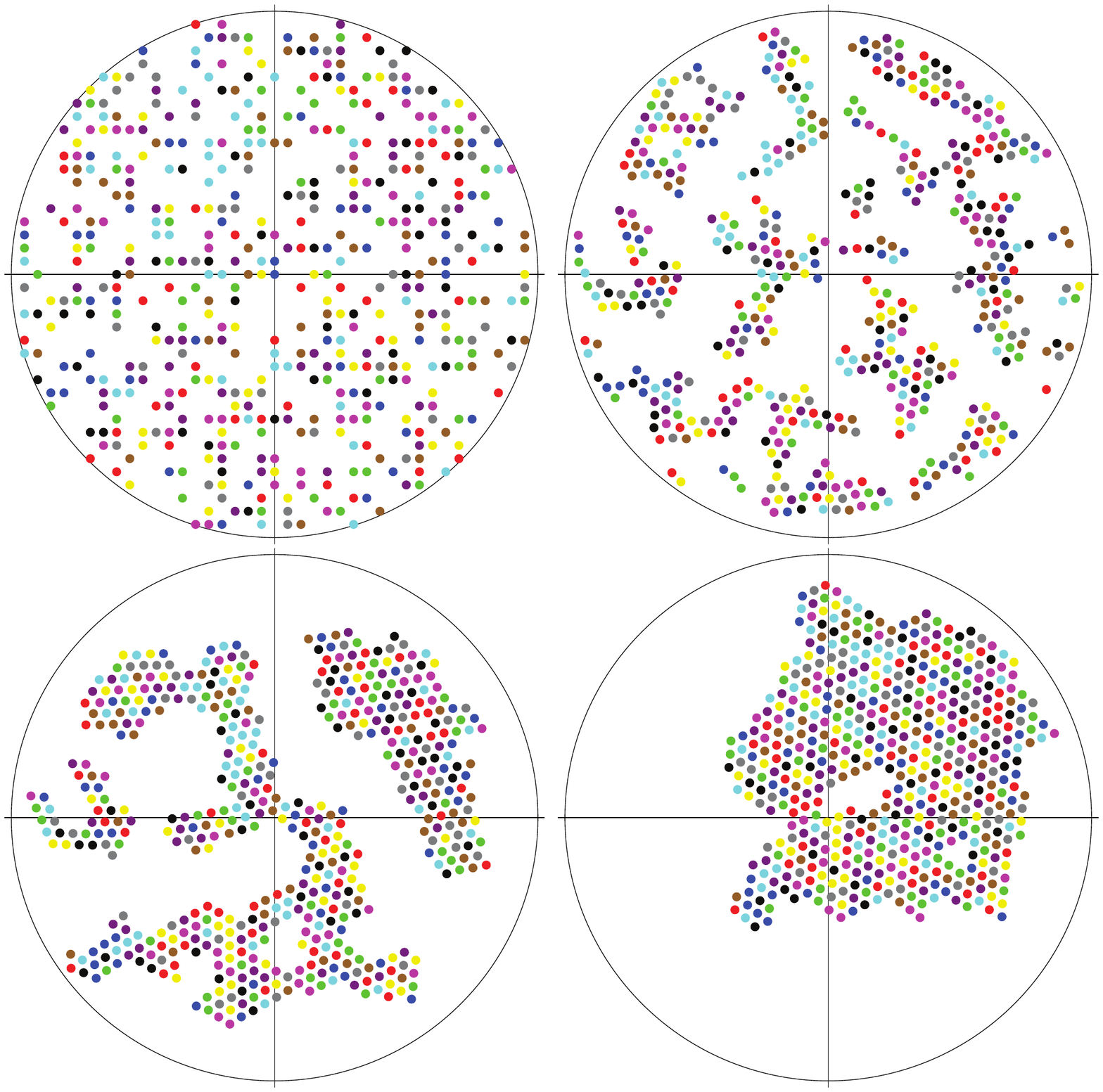}
\end{center}
\caption{\label{fig:cluster}Cluster formation in a thermostatted Lennard--Jones gas.
Simulation parameters are: $N=512,d=2,m=1,a=0.05,\varepsilon=1,\tau_{1}=0.1,\tau_{2}=0.05,T_{init}=0.5,T_{final}=0.1$}
\end{figure}

The extended system feedback force $\xi$ is fluctuating qualitatively
similar as in Fig. \ref{fig:xifig_LJCAT128.pdf}, maintaing a constant
value of the kinetic energy. Also when plotting the N-particle histograms
of the momentum distribution similar figures as previously are obtained,
see Fig. \ref{fig:histofirst_h_ANH512.pdf-1} and Fig. \ref{fig:histolast_h_ANH512.pdf}.
\\
\\
Finally we demonstrate that for sufficiently low temperatures the
thermostatted Lennard--Jones gas is forming clusters. The snapshots
were taken at the initial time and at three consecutive moments, see
Fig. \ref{fig:cluster}.\\
 
See Supplemental Material at \cite{download} for videos of the cluster
formation together with the corresponding histograms of the momenta.

\section{Outlook}

A novel type of ergostatting and thermostatting at fixed points of
the evolution of particle swarms has been presented in this paper
and shown to be viable and useful. We are convinced that various generalizations
and a new arena of exciting applications will open up. 

First we plan to check the efficiency of our thermostat~/~ergostat
for Lennard - Jones gas simulations by carefully comparing its performance
with other more conventional thermostats. In dependence of the relaxation
times $\tau_{1},\tau_{2}$ we will study - among others - variances
of the total energy, total kinetic energy and the extended system
feedback force $\xi$ \cite{neumann}. 

It seems immediately possible to formulate stochastic variants \cite{gentle,slow modes,generalized}
of our fixed point method and compare simulations with other stochastic
schemes \cite{leimkuhler efficiency}. 

A further possibility would be to extend our method to isobaric or
isothermal\textendash{}isobaric ensembles \cite{andersen,hoover pressure,constant pressure,tuckermann pressure},
where the system not only exchanges heat with the thermostat, but
also volume and work with the barostat. For a Lennard--Jones gas one
could study phase transitions and the formation of clusters. 

In a different approach we envisage to adapt our scheme to the Nosé\textendash{}
Hoover chain construction \cite{chains}, which could be interesting
especially for thermostatting small or stiff systems. 

As a final suggestion it appears interesting to examine our fixed
point method specifically in nonequilibrium conditions, where it might
be advantageous to control total energy relative to just total kinetic
energy.

\section*{Acknowledgments }

We thank William Hoover, Harald \foreignlanguage{naustrian}{Posch}
and Martin Neumann for helpful discussions. In addition, we are grateful
for financial support within the Agreement on Cooperation between
the Universities of Vienna and Zagreb.

\end{document}